\documentclass{midl} 

\usepackage{mwe} 
\usepackage[font=small,labelfont=bf, belowskip=-5pt,aboveskip=2pt]{caption}
\jmlrproceedings{}{}
\jmlrpages{}

\title[ViT-V-Net for Volumetric Medical Image Registration]{ViT-V-Net: Vision Transformer for Unsupervised Volumetric Medical Image Registration}

\midlauthor{\Name{Junyu Chen\nametag{$^{1,2}$}} \Email{jchen245@jhmi.edu}\\
\addr \vspace{-1.7em}\AND 
\Name{Yufan He\nametag{$^{1}$}} \Email{yhe35@@jhu.edu}\\
\Name{Eric C. Frey\nametag{$^{1,2}$}} \Email{efrey@jhmi.edu}\\
\Name{Ye Li\nametag{$^{1,2}$}} \Email{yli192@@jhu.edu}\\
\Name{Yong Du\nametag{$^{2}$}} \Email{duyong@jhu.edu}\\
\addr $^{1}$ Department of Electrical and Computer Engineering, Johns Hopkins University, USA\\
\addr $^{2}$ Department of Radiology and Radiological Science, Johns Hopkins Medical Institutes, USA
}

\begin{document}

\maketitle
\vspace{-12mm}

\begin{abstract}
In the last decade, convolutional neural networks (ConvNets) have dominated and achieved state-of-the-art performances in a variety medical imaging applications. However, the performances of ConvNets are still limited by lacking the understanding of long-range spatial relations in an image. The recently proposed Vision Transformer (ViT) for image classification uses a purely self-attention-based model that learns long-range spatial relations to focus on the relevant parts of an image. Nevertheless, ViT emphasizes the low-resolution features because of the consecutive downsamplings, result in a lack of detailed localization information, making it unsuitable for image registration. Recently, several ViT-based image segmentation methods have been combined with ConvNets to improve the recovery of detailed localization information. Inspired by them, we present ViT-V-Net, which bridges ViT and ConvNet to provide volumetric medical image registration. The experimental results presented here demonstrate that the proposed architecture achieves superior performance to several top-performing registration methods. Our implementation is available at \url{https://bit.ly/3bWDynR}.
\end{abstract}


\vspace{-4mm}

\section{Introduction}
\vspace{-1mm}
Deformable image registration (DIR) is fundamental for many medical image analysis tasks. It functions by of establishing spatial correspondences between points in a pair of fixed and moving images through a spatially varying deformation model. Traditionally, DIR can be performed by solving an optimization problem that maximizes the image similarity between the deformed moving and fixed images while enforcing smoothness constraints on the deformation field \cite{beg2005computing, avants2008symmetric, vercauteren2009diffeomorphic}. However, such optimization problems need to be solved for each pair of images, making those methods computationally expensive and slow in practice. Since recently, ConvNets-based registration methods \cite{de2017end, balakrishnan2018unsupervised, sokooti2017nonrigid, chen2020generating} have become a major focus of attention due to their fast computation time after training while achieving comparable accuracy to state-of-the-art methods. 

Despite ConvNets' promising performance, ConvNet architectures generally have limitations in modeling explicit long-range spatial relations (i.e., relations between two voxels that are far away from each other) present in an image due to the intrinsic locality of convolution operations \cite{chen2021transunet}. Many works have been proposed to overcoming this limitation, e.g. U-Net \cite{ronneberger2015u} (or V-Net \cite{milletari2016v}), atrous convolution (i.e, dilated convolution) \cite{yu2015multi}, and self-attention \cite{vaswani2017attention}. Recently, there has been an increasing interest in developing self-attention-based architectures due to their great success in natural language processing. Methods like non-local networks \cite{wang2018non}, detection transformer (DETR) \cite{carion2020end}, and Axial-deeplab \cite{wang2020axial} have exhibited superior performance in computer vision tasks. Dosovitskiy et al. \cite{dosovitskiy2020image} proposed Vision Transformer (ViT), a first purely self-attention-based network, and achieved state-of-the-art performance in image recognition. Subsequent to this progress, TransUnet \cite{chen2021transunet} was developed on the basis of a \textit{pre-trained} ViT for 2-dimensional (2D) medical image segmentation. However, medical imaging modalities generally produce volumetric images (i.e., 3D images), and 2D images do not fully exploit the spatial correspondences obtained from 3D volumes. Therefore, developing 3D methods is more desirable in medical image registration. In this work, we present the first study to investigate the usage of ViT for volumetric medical image registration. We propose ViT-V-Net that employs a hybrid ConvNet-Transformer architecture for self-supervised volumetric image registration. In this method, the ViT was applied to high-level features of moving and fixed images, which required the network to learn long-distance relationships between points in images. Long skip connections between encoder and decoder stages were used to retain the flow of localization information. The experimental results demonstrated that a simple swapping of the network architecture of VoxelMorph with Vit-V-Net could produce superior performance to both VoxelMorph and conventional registration methods.

\begin{figure}[t]
\floatconts
{fig_arc}
  {\caption{Method overview and network architecture of ViT-V-Net.}}
  {\includegraphics[width=0.75\textwidth]{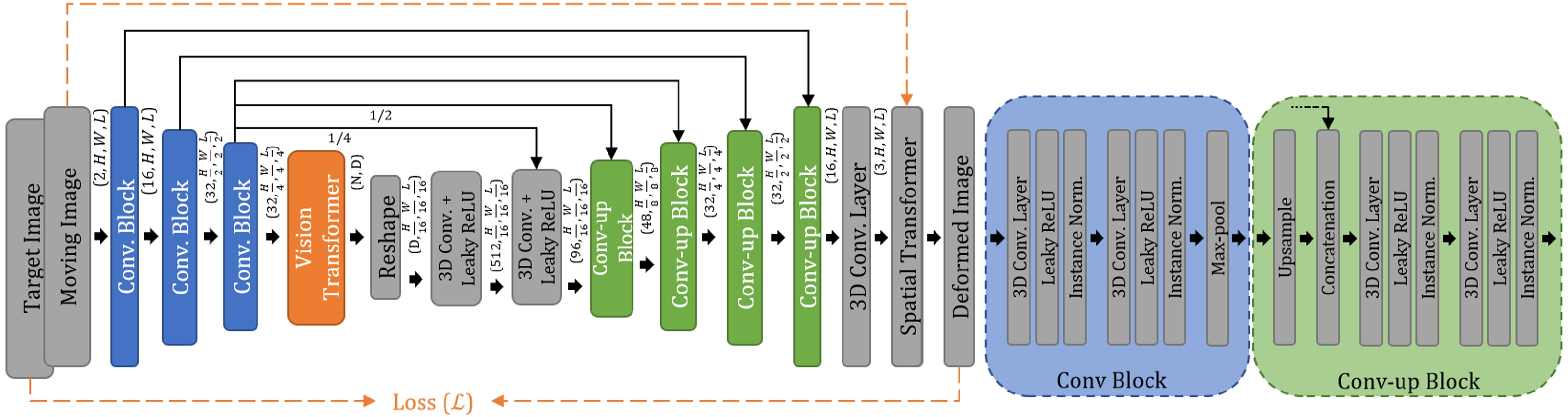}\vspace{-4mm}}
  \vspace{-4.5mm}
\end{figure}

\vspace{-5mm}
\section{Methods}
\vspace{-1mm}
Let $f\in\mathbb{R}^{H\times W\times L}$ and $m\in\mathbb{R}^{H\times W\times L}$ be fixed and moving image volumes. We assume that $f$ and $m$ are single-channel grayscale images, and they are affinely aligned. Our goal is to predict a transformation function $\mathbf{\phi}$ that warps $m$ (i.e., $m\circ\mathbf{\phi}$) to $f$, where $\mathbf{\phi}=Id+\mathbf{u}$, $\mathbf{u}$ denotes a flow field of displacement vectors, and $Id$ denotes the identity. Fig. \ref{fig_arc} presents an overview of our method. First, the deep neural network ($g_\theta$) generates $\mathbf{u}$, for the given image pair $f$ and $m$, using a set of parameters $\theta$ (i.e., $\mathbf{u} = g_\theta(f,m)$). Then, the warping (i.e., $m\circ\mathbf{\phi}$) is performed via a spatial transformation function \cite{jaderberg2015spatial}. During network training, image similarity between $m\circ\mathbf{\phi}$ and $f$ is compared, and the loss is backpropagated into the network.

\noindent\textbf{ViT-V-Net Architecture} 
Naive application of ViT to full-resolution volumetric images leads to large computational complexity. Here, instead of feeding full-resolution images directly into the ViT, the images (i.e., $f$ and $m$) were first encoded into high-level feature representations via a series of convolutional layers and max-poolings (blue boxes in Fig. \ref{fig_arc}). In the ViT (orange box), the high-level features were then separated into $N$ vectorized $P^3\times C$ patches, where $N=\frac{HWL}{P^3}$, $P$ denotes the patch size, and $C$ is the channel size. Next, the patches were mapped to a latent $D$-dimensional space using a trainable linear projection (i.e., patch embedding). Learnable position embeddings are then added to the patch embeddings to retain positional information of the patches \cite{dosovitskiy2020image}. Next, the resulting patches were fed into the Transformer encoder, which consisted of 12 alternating layers of Multihead Self-Attention (MSA) and Multi-Layer Perceptron (MLP) blocks \cite{vaswani2017attention} (see Appendix \ref{ViT_overview} for details of ViT). Finally, the output from ViT was reshaped and then decoded using a V-Net style decoder. Notice that long skip connections between the encoder and decoder were also used. The network's final output is a dense displacement field, which was then used in the spatial transformer for warping $m$.

\noindent\textbf{Loss Functions} 
The image similarity measurement used in this study was mean squared error (MSE), along with a diffusion regularizer controlled by a weighting parameter $\lambda$ for imposing smoothness in the displacement field $\mathbf{u}$ (see Appendix \ref{Loss} for formulation).

\begin{table}[h]
\vspace{-1mm}
\centering
\scriptsize
    \begin{tabular}{ c | c c c c c c}
 \hline
 & Affine only & NiftyReg & SyN & VoxelMoprh-1 & VoxelMoprh-2 & ViT-V-Net\\
 \hline
 \textbf{Dice} & 0.569$\pm$0.171 & 0.713$\pm$0.134 & 0.688$\pm$0.140 & 0.707$\pm$0.137 & 0.711$\pm$0.135 & \textbf{0.726$\pm$0.130}\\
 
 \hline
\end{tabular}
\captionsetup{justification=centering}
\caption{Overall Dice comparisons between the proposed method and the others. Detailed performance on various anatomical structures are shown in Fig. \ref{fig_vit_res} of Appendix \ref{add_res}.}\label{table_r}
\end{table}
\vspace{-8mm}

\begin{figure}[t]
\floatconts
{fig_quali}
  {\caption{Registration results of a MR coronal slice. Additional results are shown in Appendix \ref{add_res}.}}
  {\includegraphics[width=0.7\textwidth]{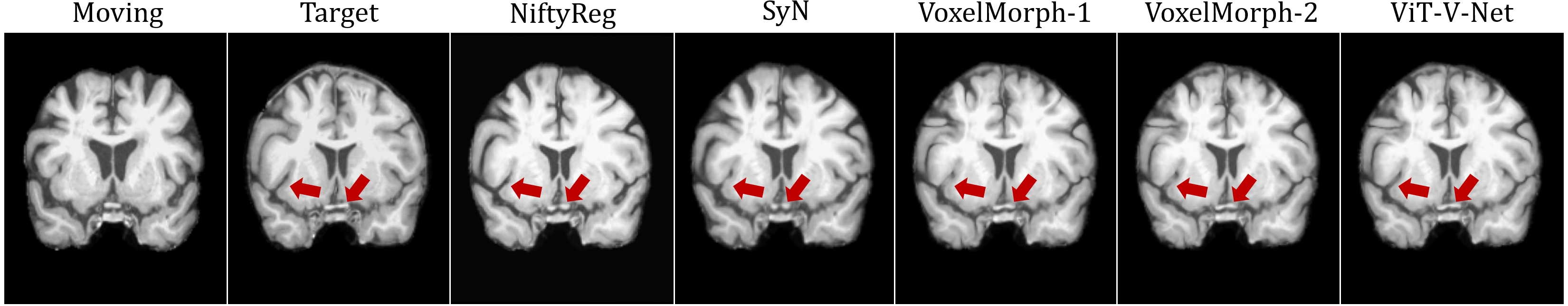}\vspace{-4mm}}
\vspace{-4mm}
\end{figure}

\section{Results and Conclusions}
\vspace{-1.5mm}
We demonstrate our method on the task of brain MRI registration. We used an in-house dataset that consists of 260 T1–weighted brain MRI scans. The dataset was split into 182, 26, and 52 (7:1:2) volumes for training, validation, and test sets. Each image volume was randomly matched to two other volumes to form four pairs of $f$ and $m$, resulting in 768, 104, and 208 image pairs. Standard pre-processing steps for structural brain MRI, including skull stripping, resampling, and affine transformation were performed using FreeSurfer \cite{fischl2012freesurfer}. Then, the resulting volumes were cropped to an equal size of $160\times192\times 224$. Label maps including 29 anatomical structures were obtained using FreeSurfer for evaluation. The proposed method was compared in terms of Dice score \cite{dice1945measures} to Symmetric Normalization (SyN)\footnote{Implementation of SyN was obtained from \url{https://github.com/ANTsX/ANTsPy}} \cite{avants2008symmetric}, NiftyReg\footnote{Implementation of NiftyReg was obtained from \url{https://www.ucl.ac.uk/medical-image-computing}} \cite{modat2010fast}, and a learning-based method, VoxelMorph\footnote{Implementation of VoxelMorph was obtained from \url{http://voxelmorph.csail.mit.edu}}-1 and -2 \cite{balakrishnan2018unsupervised}. The regularization parameter, $\lambda$, was set to be 0.02, which was reported in \cite{balakrishnan2018unsupervised} as an optimal value for VoxelMorph. The method was implemented using PyTorch \cite{NEURIPS2019_9015}. Detailed hyperparameter settings for training are shown in Appendix \ref{hyper}. Qualitative results, and Dice scores are shown in Table \ref{table_r} and Fig. \ref{fig_quali}. As visible from the results, the proposed ViT-V-Net yielded a significant gain of $>0.1$ in Dice performance ($p$-values are shown in Table. \ref{table_res}) compared to the others. We also noticed that ViT-V-Net reached lower loss values and had higher validation Dice scores during training (see Fig. \ref{fig_curve} in Appendix \ref{add_res}). In conclusion, the proposed ViT-based architecture achieved superior performance than the top-performing registration methods, demonstrating the effectiveness of ViT-V-Net.

\midlacknowledgments{This work was supported by a grant from the National Cancer Institute, U01-CA140204. The views expressed in written conference materials or publications and by speakers and moderators do not necessarily reflect the official policies of the NIH; nor does mention by trade names, commercial practices, or organizations imply endorsement by the U.S. Government.}

\bibliography{midl-samplebibliography}

\newpage
\appendix
\section{Overview of Vision Transformer}
A detailed description of ViT can be found in \cite{dosovitskiy2020image, vaswani2017attention, chen2021transunet}.
\label{ViT_overview}
\begin{figure}[!h]
\floatconts
{fig_vit}
  {\caption{Model overview of the Vision Transformer.}}
  {\includegraphics[width=0.35\textwidth]{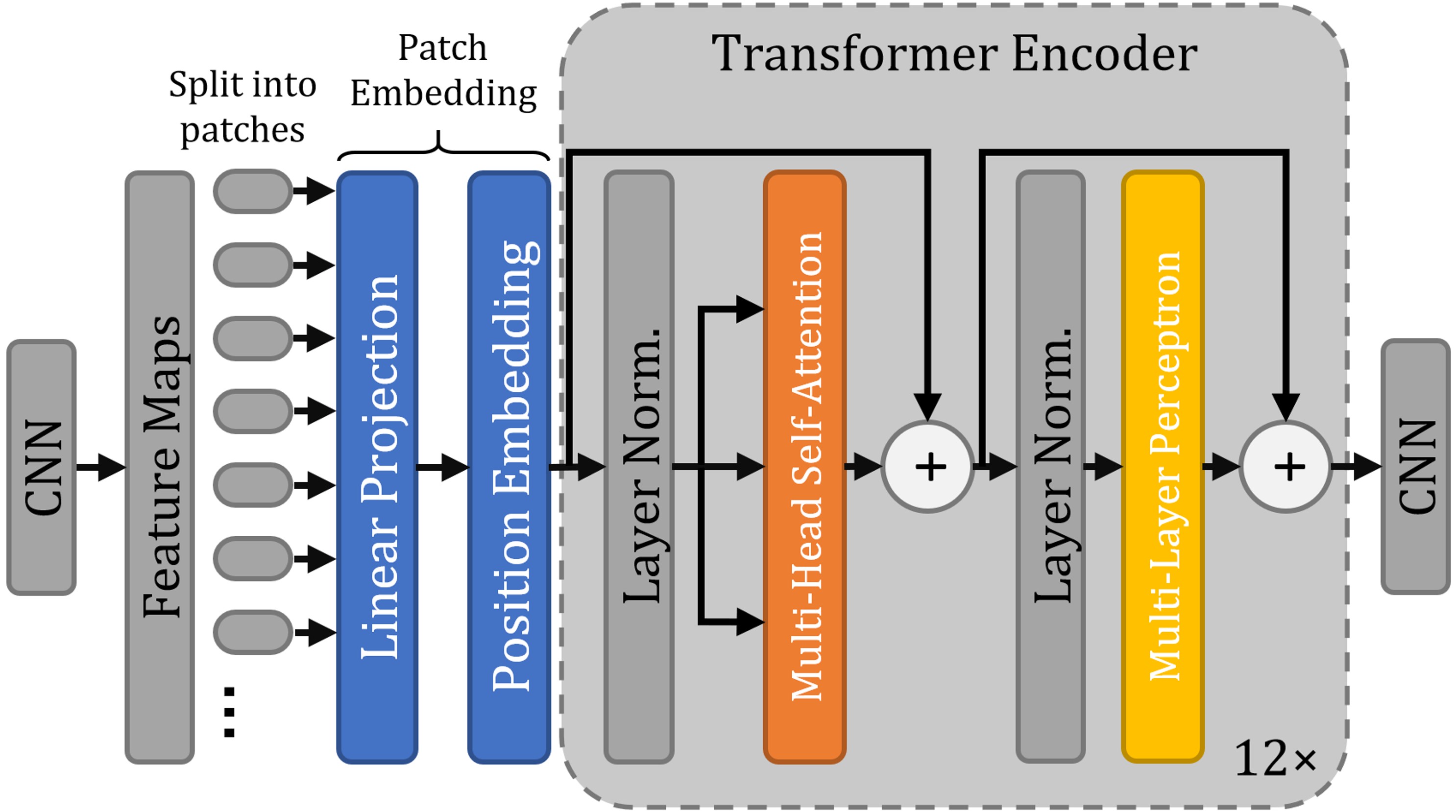}}
\end{figure}

\noindent\textbf{Patch Embedding}
Let $x_p^i$ be the $i^{th}$ vectorized patch, where $i\in\{1,...,N\}$. The patches were first encoded into a latent $D$-dimensional space using a trainable linear projection (realized via a convolutional layer). Then, learnable position embeddings were added to retain positional information:
\begin{equation}
    \mathbf{z}_0 = [x_p^1\mathbf{E};x_p^2\mathbf{E};...;x_p^N\mathbf{E}]+\mathbf{E}_{pos},
\end{equation}
where $\mathbf{E}\in\mathbb{R}^{P^3C\times D}$ denotes the patch embedding projection and $\mathbf{E}_{pos}\in\mathbb{R}^{N\times D}$ represents the positional embedding matrix. Next, the output $\mathbf{z}_0$ was fed into consecutive blocks of the Transformer encoder.

\noindent\textbf{Transformer encoder} The Transformer encoder consists of 12 blocks of MSA and MLP layers \cite{vaswani2017attention}. A layer normalization (LN) was applied before each MSA and MLP layer. The output of $\ell^{th}$ Transformer encoder can be written as:
\begin{equation}
\begin{split}
    \mathbf{z}_\ell'&= \text{MSA}(\text{LN}(\mathbf{z}_{\ell-1})) + \mathbf{z}_{\ell-1}\\
    \mathbf{z}_\ell&= \text{MLP}(\text{LN}(\mathbf{z}_\ell')) + \mathbf{z}_\ell',
\end{split}
\end{equation}
where $\mathbf{z}_\ell$ denotes the encoded image representation.

\section{Loss Functions}
\label{Loss}
The loss function used for training the proposed network can be written as:
\begin{equation}
    \mathcal{L}(f, m, \mathbf{\phi}) = \mathcal{L}_{MSE}(f, m, \mathbf{\phi}) + \lambda\mathcal{L}_{diffusion}(\mathbf{\phi}),
\end{equation}
where $\lambda$ is a regularization parameter, $f$ and $m$ are, respectively, the fixed and moving image, and $\mathcal{\phi}$ represents the deformation field.

\noindent\textbf{Image Similarity Measurement}
The mean squared error (MSE) between the deformed moving image and fixed image was used as the loss function. It is defined as:
\begin{equation}
    \mathcal{L}_{MSE}(f, m, \mathbf{\phi}) =\frac{1}{\Omega}\sum_{p\in\omega}\left[f(p)-m\circ\mathbf{\phi}(p)\right]^2,
\end{equation}
where $\Omega$ denotes the image domain.

\noindent\textbf{Deformation Field Regularization}
To enforce smoothness in the deformation field, a diffusion regularizer was used. It is defined as:
\begin{equation}
    \mathcal{L}_{diffusion}(\mathbf{\phi}) =\sum_{p\in\omega}\Vert\nabla\mathbf{u}(p)\Vert^2,
\end{equation}
where $\mathbf{u}$ the displacement field, which is the output of the network.

\section{Hyperparameters Settings}
\label{hyper}
\begin{table}[!hbp]
\centering
\footnotesize
    \begin{tabular}{ c | c c c}
 \hline
  & VoxelMoprh-1 & VoxelMoprh-2 & ViT-V-Net\\
 \hline
 Optimizer & ADAM & ADAM & ADAM \\
 \hline
 Learning rate& $1e^{-4}$ & $1e^{-4}$ & $1e^{-4}$\\
  \hline
 Learning rate decay & Polynomial (0.9) & Polynomial (0.9) & Polynomial (0.9)\\
  \hline
Dropout & $0.0$ & $0.0$ & $\mathbf{0.1}$\\
  \hline
 Epochs & $500$ & $500$ & $500$\\
 \hline
 Batch size  & $2$ & $2$ & $2$\\
 \hline
 Loss function & MSE & MSE & MSE \\
 \hline
 Regularizer & Diffusion & Diffusion & Diffusion \\
 \hline
 Regularization parameter ($\lambda$)  & $0.02$ & $0.02$ & $0.02$\\
 \hline
 Data augmentation  & Random flipping & Random flipping & Random flipping\\
 \hline
 ViT patch size ($P$)  & - & - & 8\\
 \hline
 ViT latent vector size ($D$)  & - & - & 252\\
 \hline
 GPU memory used during training  & 17.320 GiB & 19.579 GiB & 18.511 GiB\\
 \hline
\end{tabular}
\captionsetup{justification=centering}
\caption{Training setups for the learning-based models. All models were trained using the same optimizer (ADAM \cite{kingma2014adam}) and training hyperparameters, except the dropout rate was set to be 0.1 in linear layers of ViT-V-Net. The models were trained and tested on a PC with an AMD Ryzen 9 3900X CPU, an NVIDIA Titan RTX GPU, and an NVIDIA RTX 3090 GPU, where both GPUs have 24 GiB memory. Each model took about 3 days to train on a single GPU.}\label{table_hyper}
\end{table}

\begin{table}[!hbp]
\centering
\scriptsize
    \begin{tabular}{ c | c c c c}
 \hline
 & Cost fuction & Regularizer  & Regularization parameter & Number of iteration\\
 \hline
 NiftyReg & SSD & Bending energy (default)& 0.0002 & 300, 300, 300 (default)\\
 \hline
 SyN & MSQ & Gaussian (default) & 3 (default) & 40, 20, 0 (default)\\
 \hline
\end{tabular}
\captionsetup{justification=centering}
\caption{Hyperparameter settings for NiftyReg and SyN, where SSD stands for the sum of squared difference and MSQ stands for the mean squared difference. We chose the regularization parameter for NiftyReg to be 0.0002, because the default value, 0.005, led to over-smoothed suboptimal deformations.}\label{table_hyperconven}
\end{table}

\section{Additional Results}
\label{add_res}
\begin{table}[!hbp]
\centering
\scriptsize
    \begin{tabular}{ c | c c c c c}
 \hline
 & NiftyReg & SyN  & VoxelMorph-1 & VoxelMorph-2 & ViT-V-Net\\
 \hline
 Dice & 0.713$\pm$0.134 & 0.688$\pm$0.140 & 0.707$\pm$0.137 & 0.711$\pm$0.135 & 0.726$\pm$0.130\\
 \hline
 \% of $\vert J_{\phi}\vert\leq0$ & 0.225$\pm$0.165& 0.118$\pm$0.084 & 0.375$\pm$0.098 & 0.414$\pm$0.084 & 0.381$\pm$0.102\\
 \hline
 Time (Sec) & 113 & 15.257 & 0.002 & 0.002 & 0.002\\
 \hline
\end{tabular}
\captionsetup{justification=centering}
\caption{Quantitative comparisons of Dice score, percentage of voxels with a non-positive Jacobian determinant (i.e., folded voxels), and computational time for different methods. Note that NiftyReg and SyN were applied using the CPUs, while the learning-based methods, VoxelMorph and ViT-V-Net, were implemented on GPU.}\label{table_res}
\end{table}

\begin{table}[!htp]
\centering
\small
    \begin{tabular}{ c | c c c c c}
 \hline
 & Affine & NiftyReg & SyN & VoxelMorph-1 & VoxelMorph-2\\
 \hline
 ViT-V-Net & $\ll1e^{-5}$ & $4.102e^{-3}$ & $\ll1e^{-5}$ & $1.536e^{-5}$ & $1.601e^{-3}$\\
 \hline
\end{tabular}
\captionsetup{justification=centering}
\caption{$p$-values computed using the paired t-test on the Dice scores between the proposed ViT-V-Net and other registration methods.}\label{table_ttest}
\end{table}

\newpage
\begin{figure}[!htp]
\floatconts
{fig_curve}
  {\caption{Training loss value and validation Dice score per epoch. The proposed ViT-V-Net exhibits lower loss values and higher Dice scores during training.}}
  {\includegraphics[width=1\textwidth]{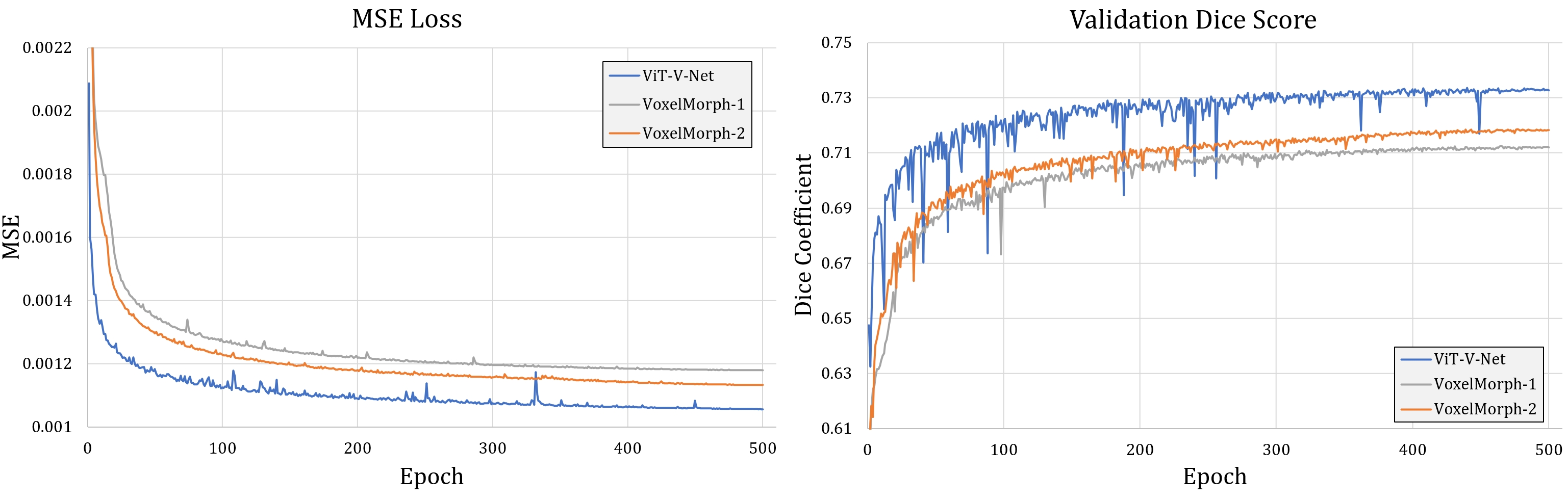}}
\end{figure}

\begin{figure}[!hbp]
\floatconts
{fig_vit_res}
  {\caption{Boxplots of Dice scores for various anatomical structures obtained using different registration methods. Dice scores of the left and right brain hemispheres were averaged into a single score. Orange triangles denote the means.}}
  {\includegraphics[width=1\textwidth]{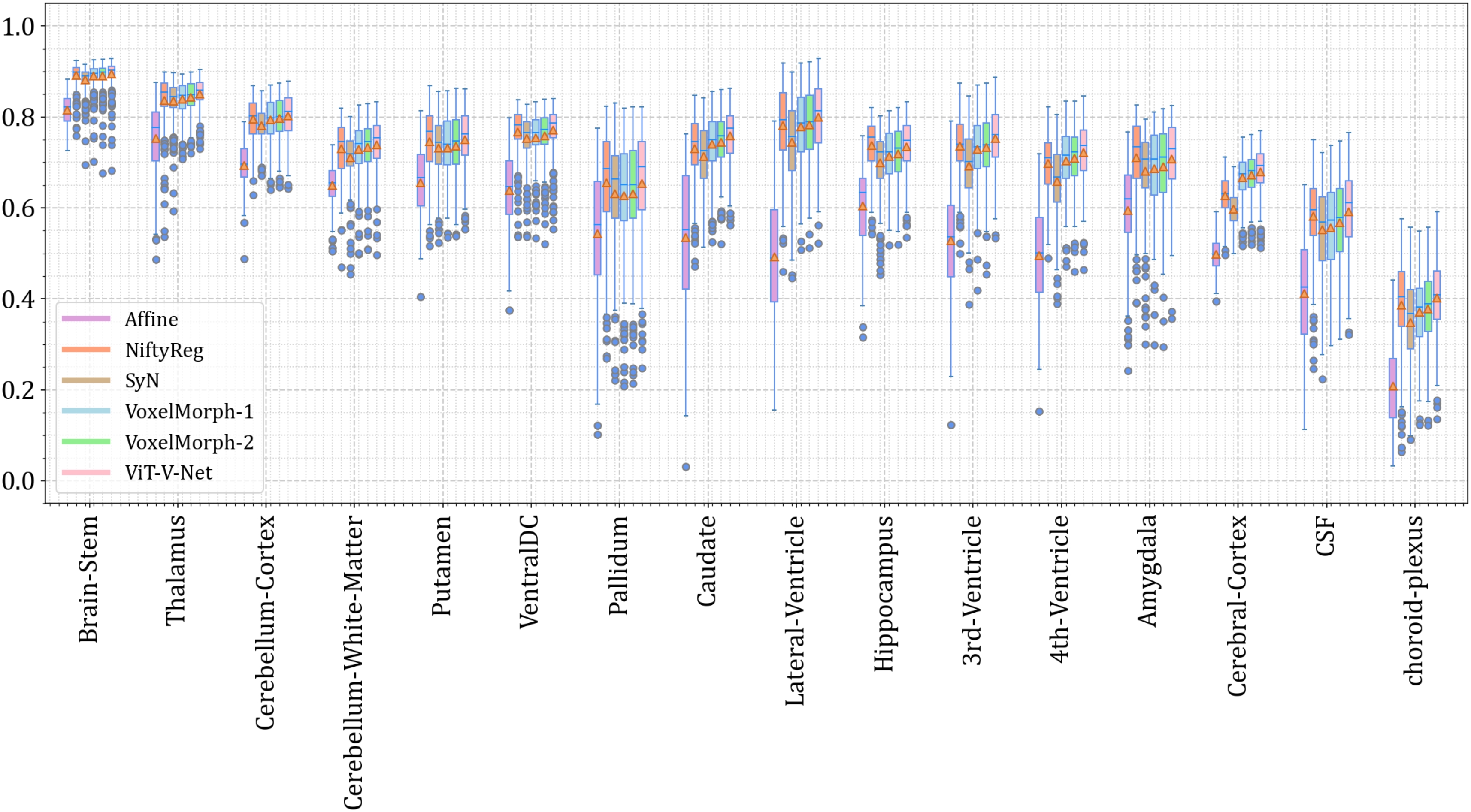}}
\end{figure}
\newpage
\vspace{-4mm}
\begin{figure}[!htp]
\floatconts
{fig_vit_res}
  {\vspace{-4mm}\caption{Additional qualitative results generated by different registration methods. The rows in the top panel (of two rows separated by a dashed line) show, respectively, moving and fixed images. The other five panels exhibit deformed images and their corresponding displacement fields produced by different methods. The colored images were created by first clamping the displacement values to a range of $[-10, 10]$ and then mapping each spatial dimension to each of the RGB color channels.}}
  {\includegraphics[trim={1cm 2cm 2cm 2.2cm},clip, width=0.87\textwidth]{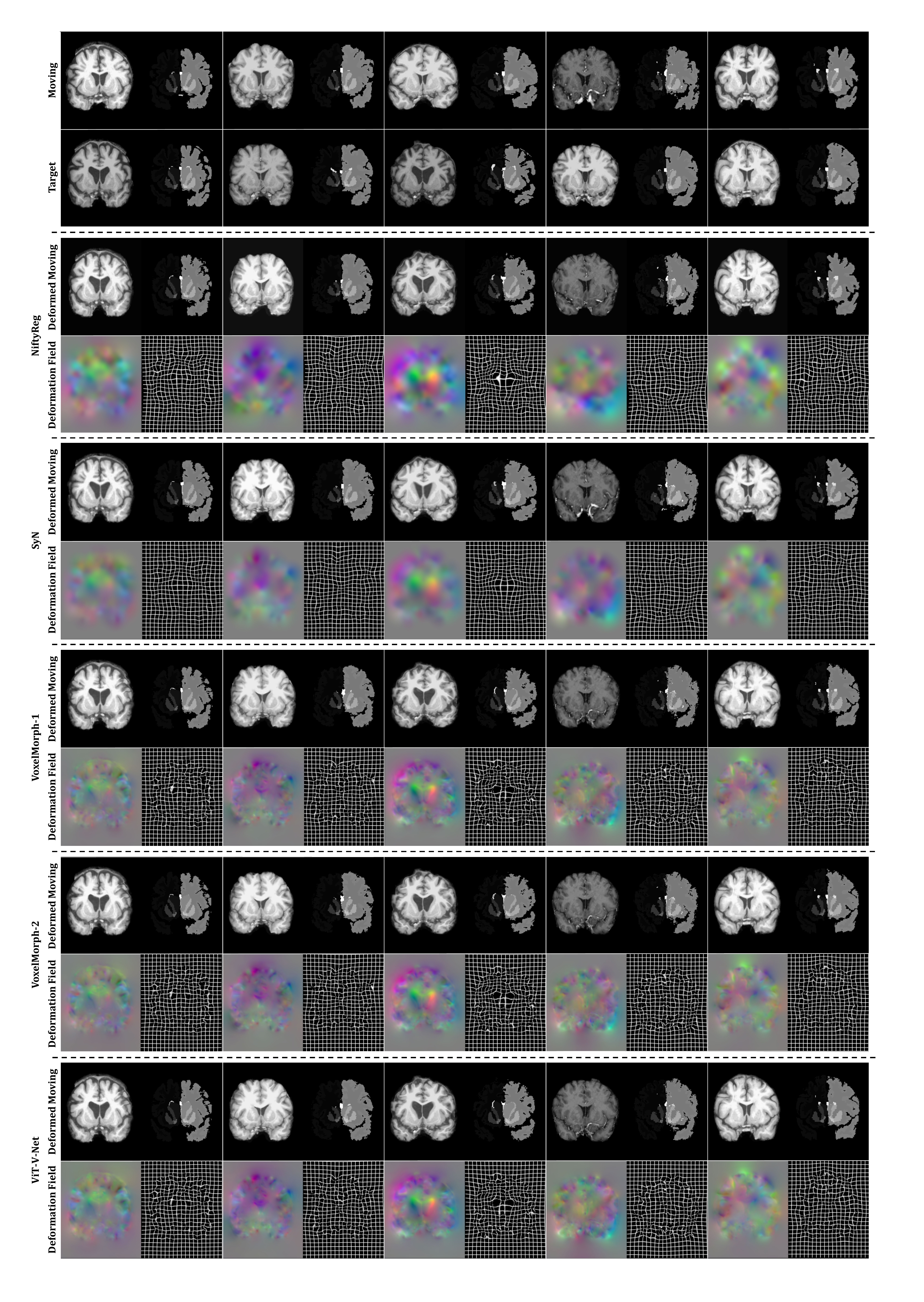}}
  
\end{figure}

\let\cleardoublepage\clearpage
\end{document}